# Linear analysis of the 4-step Carré phase shifting algorithm: spectrum, signal-to-noise ratio, and harmonics response

**Manuel Servin and Adonai Gonzalez**

*Centro de Investigaciones en Optica A.C., Loma del Bosque 115, Leon Guanajuato, 37000 Mexico.*
*First author's e-mail: mservin@cio.org*

**Abstract:** We analyze the nonlinear Carré 4-steps algorithm including its frequency response, signal-to-noise ratio, and harmonics rejection using linear systems theory. At first sight the previous statement as well as the title of this paper seems paradoxical. How can we analyze the 4-step non-linear Carré Phase Shifting Algorithm (PSA) using linear system theory? The short answer is that the non-linear Carré algorithm may be decomposed into two building blocks. The first block is a tunable linear 4-step PSA, and the second one is a non-linear phase-step estimator. Although this fact is well known from the derivation of the Carré algorithm, nobody has properly exploited it. In other words, to this day, we do not have explicit mathematical formulae for a) the spectrum, b) the harmonics rejection, and c) the signal-to-noise ratio of the non-linear Carré algorithm. These are the properties of the Carré's PSA that we show here with novel and explicit mathematical formulae.

## 1. Introduction

The non-linear Carré algorithm [1] was the first and best known phase-shifting algorithm (PSA) which allows one to recover the modulating phase from a set of 4 phase-shifted interferograms. The advantage of the Carré PSA is that it allows demodulating the phase of 4 interferograms without explicitly knowing the constant phase-step among them [2-6]. We may call this non-linear algorithms self-tuning PSA, given that it finds its phase-step and its modulating phase simultaneously [5-10]. Despite that the Carré 4-step PSA is a non-linear system, this PSA may be regarded as composed of a linear PSA plus a non-linear phase-step estimator [2,5,6]. Although this fact has been known for a long time [2-10], it has not been properly analyzed. As far as we know, the only study which has exploited this fact, but failed is Malacara et al. book [2]. The herein proposed new arrangement of the Carré algorithm allows its definitive understanding. This novel arrangement permits to decouple the linear and the mom-linear parts of the Carré algorithm offering new explicit formulae for: a) its frequency spectrum, b) its signal-to-noise ratio (*S/N*), and c) its harmonics rejection [10].

For instance as we show here, the non-linear part of the Carré algorithm which estimates its phase-step, remains unchanged for any time or phase shift of the algorithm. Also the non-linear part estimation of the phase-step does not play any role in the spectra of the algorithm.

In summary, to this day, there is no published work that gives explicit formulae for: a) the spectrum, b) the harmonic rejection, and c) the (*S/N*) of the Carré algorithm. That is because everybody (except for the erroneous analysis made in [2]) has focused in the non-linear version of it which can not be analyzed using linear system theory.

Because the non-linear Carré algorithm has been around for about 46 years, we may naively think that the results herein presented are "well known facts"; we will do our best to avoid this trap. To that end we apologize if this paper sounds "too defensive", heavily emphasizing each new result, and new explicit mathematical formulae.

## 2. Well known, published facts about the non-linear 4-step Carré PSA

In this section we review the well known, published facts about the 4-step Carré PSA [1-10].
The standard mathematical model of a temporal phase-shifted interferogram is,

$$I(x,y,t) = a(x,y) + b(x,y)\cos[\varphi(x,y) + \omega_0 t], \qquad t \in \mathbb{R}. \qquad (1)$$

Where $a(x,y)$ is the background illumination, $b(x,y)$ is the contrast of the fringes, $\varphi(x,y)$ is the modulating phase to be estimated, $\omega_0$ is the temporal carrier frequency in radians per interferogram, and finally $t$ is a real number. Time $t$ is only formal; one is only interested in total phase-shift. In particular the 4 temporal steps used for the Carré algorithm are [5],

$$\begin{aligned}
I(x,y,-3/2) &= a(x,y) + b(x,y)\cos[\varphi(x,y) - 3\omega_0/2], \\
I(x,y,-1/2) &= a(x,y) + b(x,y)\cos[\varphi(x,y) - \omega_0/2], \\
I(x,y,1/2) &= a(x,y) + b(x,y)\cos[\varphi(x,y) + \omega_0/2], \\
I(x,y,3/2) &= a(x,y) + b(x,y)\cos[\varphi(x,y) + 3\omega_0/2].
\end{aligned} \qquad (2)$$

The phase steps are $\{-3\omega_0/2, -\omega_0/2, \omega_0/2, 3\omega_0/2\}$. These 4 equations have 4 unknowns: $a(x,y)$, $b(x,y)$, $\varphi(x,y)$, and $\omega_0$. So we can solve for $\omega_0$, and $\varphi(x,y)$ as a function of the data: $s_1 = I(x,y,-3/2)$, $s_2 = I(x,y,-1/2)$, $s_3 = I(x,y,1/2)$, $s_4 = I(x,y,3/2)$ obtaining [2,3]:

$$\tan[\hat{\varphi}(x,y)] = \frac{s_1 + s_2 - s_3 - s_4}{-s_1 + s_2 + s_3 - s_4}\tan\left[\frac{\omega_0(x,y)}{2}\right], \qquad \hat{\varphi}(x,y) \in (-\pi, \pi). \qquad (3)$$

Where $\hat{\varphi}(x,y)$ is the estimated phase which may differ a bit from the true modulating phase $\varphi(x,y)$. Note that the PSA in Eq. (3) has a linear *non-tunable* PSA, multiplied by the tangent of $\omega_0(x,y)/2$. In Eq. (3), the linear non-tunable part is a 4-step least-squares PSA, tuned at the fixed frequency $\omega_0(x,y) = \pi/2$ [2]. From Eq. (2) one estimates $\omega_0(x,y)$ as,

$$\tan\left[\frac{\omega_0(x,y)}{2}\right] = \sqrt{\frac{3(s_2 - s_3) - s_1 + s_4}{s_1 + s_2 - s_3 - s_4}}, \qquad \omega_0(x,y) \in (0, \pi) \qquad (4)$$

It is common to combine these two results Eqs. (3-4) into a single non-linear equation obtaining the well known 4-step Carré PSA [2,3,5,6],

$$\tan[\hat{\varphi}(x,y)] = \frac{\sqrt{3(s_2 - s_3)^2 - (s_1 - s_4)^2 + 2(s_1 - s_4)(s_2 - s_3)}}{-s_1 + s_2 + s_3 - s_4}, \qquad \hat{\varphi}(x,y) \in (0, \pi). \qquad (5)$$

The positive sign of the square root in Eq. (5) forces the estimated phase $\hat{\varphi}(x,y)$ to be wrapped modulo $\pi$. To transform the phase modulo $\pi$ into a usable modulo $2\pi$ one, Creath [5,6] gave an algorithmic criteria to collocate $\hat{\varphi}(x,y)$ (Eq.(5)) in the right quadrant of the unit circle. By 2007 Schreiber and Bruning in [3] gave again the same receipt that Creath [5,6] provided, so they [3] also encourages the use of the non-linear formula in Eq. (5).

Malacara et. al. [2] analyzed the linear *non-tunable* part of Eq. (3) which reads:

$$\frac{s_1 + s_2 - s_3 - s_4}{-s_1 + s_2 + s_3 - s_4}. \qquad (6)$$

They used the Freischlad and Koliopoulos (F&K) [4] spectral theory (Eqs. (6.171) and (6.172) in [2]) to analyze it. But this analysis is incomplete because it cover *only* the special case when $\omega_0$ is exactly $\pi/2$ or $\tan(\omega_0/2) = 1$. Afterwards Malacara et. al. [2] tried to analyze the full tunable version in Eq. (3), but failed because they assigned the tangent function to the numerator of Eq. (3); giving an erroneous F&K spectra (Eqs. (6.177) and (6.178) in [2]).

### 3. Some erroneous "well known" properties about the Carré algorithm

Since 1966 [1] erroneous facts about the Carré algorithm have been published, including:
  a) The Carré PSA has its best signal-to-noise (*S/N*) ratio at $\omega_0 = 65$ degrees [8].
  b) The Carré PSA has its best (*S/N*) ratio at $\omega_0 = 110$ degrees [2,4].
  c) The Carré PSA is insensitive to all even harmonics [2].

This paper corrects these erroneous published properties.

### 4. Linear analysis of the 4-step Carré algorithm

The first small, but useful step that we propose, is to separate the tangent function in Eq. (3), into its sine and cosine functions, as:

$$\tan[\hat{\varphi}(x,y)] = \frac{(s_1 + s_2 - s_3 - s_4)\sin(\omega_0/2)}{(-s_1 + s_2 + s_3 - s_4)\cos(\omega_0/2)}, \qquad \hat{\varphi}(x,y) \in (-\pi, \pi). \qquad (7)$$

And then find the impulse filter response associated with this algorithm [11,12,13] as:

$$h(t) = \left[-\delta(t+\frac{3}{2}) + \delta(t+\frac{1}{2}) + \delta(t-\frac{1}{2}) - \delta(t-\frac{3}{2})\right]\cos\left(\frac{\omega_0}{2}\right)$$
$$+ i\left[\delta(t+\frac{3}{2}) + \delta(t+\frac{1}{2}) - \delta(t-\frac{1}{2}) - \delta(t-\frac{3}{2})\right]\sin\left(\frac{\omega_0}{2}\right), \qquad (8)$$

where $i = \sqrt{-1}$, and $\omega_0(x,y)$ is given by Eq. (4). The system in Eq. (8) differs from Eq. (6) for all $\omega_0$ except for $\pi/2$.

The Fourier transform $H(\omega) = F[h(t)]$ (the FTF [11,12,13]) of this filter is,

$$H(\omega) = \left[-\exp(\frac{3i\omega}{2}) + \exp(\frac{i\omega}{2}) + \exp(-\frac{i\omega}{2}) - \exp(-\frac{3i\omega}{2})\right]\cos\left(\frac{\omega_0}{2}\right)$$
$$+ i\left[\exp(\frac{3i\omega}{2}) + \exp(\frac{i\omega}{2}) - \exp(-\frac{i\omega}{2}) - \exp(-\frac{3i\omega}{2})\right]\sin\left(\frac{\omega_0}{2}\right) \qquad (9)$$

After some algebra one obtains [11].

$$H(\omega) = e^{i\frac{3\omega}{2}} e^{i\frac{\omega_0}{2}} \left[1 - e^{-i\omega}\right]\left[1 - e^{-i(\omega+\omega_0)}\right]\left[1 - e^{-i(\omega+\pi)}\right]. \qquad (10)$$

The phase factor $\exp(i3\omega/2)$ is related with the displacement of 3/2 pixels of the data (Eq. (2)). The phase factor $\exp(i\omega_0/2)$ appears because the data carrier appears as multiples of $\omega_0/2$. Apart from these two *irrelevant* phase factors, Eq. (10) is equal (have the same FTF $|H(\omega)|$ and phase demodulation properties) than the following one,

$$H_2(\omega) = \left[1 - e^{-i\omega}\right]\left[1 - e^{-i(\omega+\omega_0)}\right]\left[1 - e^{-i(\omega+\pi)}\right]. \qquad (11)$$

This result has never been published, neither in the form of a F&K spectra nor using the Surrel's *x*-polynomials [12]. The Frequency Transfer Function in Eq.(11) may also be expressed using Surrel's characteristic *x*-polynomials [12] (with $x = e^{-i\omega}$) as:

$$P(x) = (1-x)(1 - e^{-i\omega_0} x)(1+x). \qquad (12)$$

All these spectra Eq. (10), Eq. (11), and Eq. (12) are equivalent and have the same carrier $\omega_0$ give by Eq. (4). From the spectrum in Eq. (11) we see that it has a movable zero at $\omega_0$. This movable zero is the one that give the *singular* tunable property of the Carré algorithm. Note that the useful form of the equivalent spectra in Eqs. (10,11,12) have never been published. As the next sections show, this spectrum is the key to understand all the properties of the Carré algorithm. From now on we will only refer to the spectral form in Eq. (11).

## 5. Frequency response of the non-linear Carré PSA

For the reader's convenience, here we repeat Eq. (11),

$$H_2(\omega) = \left[1 - e^{-i\omega}\right]\left[1 - e^{-i(\omega+\omega_0)}\right]\left[1 - e^{-i(\omega+\pi)}\right]. \qquad (11)$$

Eq. (11) shows that, the first zero remains fixed at the origin $\omega = 0$, the third zero is also fixed but at $\omega = \pm\pi$, *and the middle one is at $\omega = -\omega_0$; it is tunable*. The interesting thing about the Carré PSA resides in the tunable zero at $\omega = -\omega_0$. This zero is the only "movable zero" of this algorithm. Varying this zero within $\omega_0 \in (-\pi, 0)$ one generates a continuous family of 4-step linear quadrature filters which rejects the frequencies at $\omega = \{0, -\omega_0, \pm\pi\}$. Figure 1 shows the magnitude of the FTF $|H_2(\omega)|$ (Eq. (11)) for three different tuning frequencies: $\omega_0 = 0.25\pi$, $\omega_0 = 0.5\pi$, and $\omega_0 = 0.75\pi$.

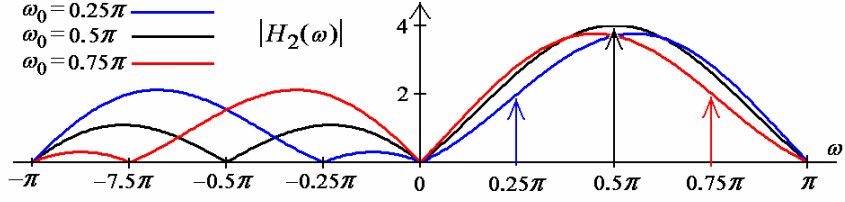

Figure 1. Magnitude of the FTF of the filter in Eq. (11) for three different carrier frequencies $\omega_0$ at $\{0.25\pi, 0.5\pi, 0.75\pi\}$

We see that the maximal signal amplitude of the Carré algorithm is obtained when it is tuned at $\omega_0 = 0.5\pi$ (the black plot) and its maximum value is $|H_2(0.5\pi)|_{Max} = 4$. The form of the spectra in Fig. 1, tuned at different carriers shows the adaptability of the Carré algorithm to the incoming interferometric signal's carrier $\omega_0$. The explicit knowledge of this spectral behaviour, although implicit in the Carré algorithm has never been published, and it is very useful to properly understand this algorithm.

## 6. Signal-to-noise (S/N) power ratio gain $G(\omega_0)$ of the non-linear Carré PSA

We now turn our attention to the signal-to-noise (*S/N*) power ratio of the Carré algorithm. The signal-to-noise power ratio $G(\omega_0)$ of a PSA, with data corrupted by white additive noise may be expressed as,

$$\left(\frac{S}{N}\right)_{\text{Output Signal}} = G(\omega_0)\left(\frac{S}{N}\right)_{\text{Input Signal}}. \quad (13)$$

The signal to noise ratio gain $G(\omega_0)$ is given by [13],

$$G(\omega_0) = \frac{|H_2(\omega_0)|^2}{\frac{1}{2\pi}\left(\int_{-\pi}^{\pi}|H_2(\omega)|^2 d\omega\right)}. \quad (14)$$

When $G(\omega_0) = 1.0$, the output analytical signal has the same (*S/N*) power as the interferograms. When $G(\omega_0) > 1.0$ the output signal has a higher (*S/N*) ratio than the interferometric signal; the standard case. Finally when $G(\omega_0) < 1.0$, the analytical output signal (the PSA's output) has a reduced (*S/N*) than the input signal; this situation is not desired. Using Eq. (14), the (*S/N*) ratio gain for the Carré algorithm is shows in Figure 2. One may see that the maximum (*S/N*) power ratio gain is 4 and it is achieved for a carrier frequency of $\omega_0 = \pi/2$.

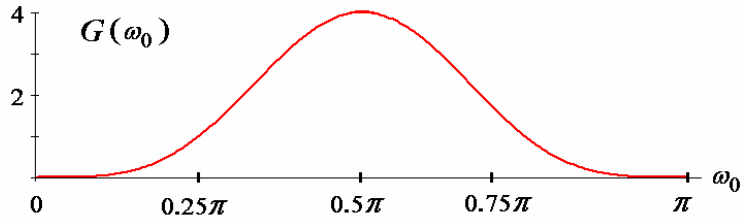

Figure 2. Signal-to-noise power ratio $G(\omega_0)$ of the interferometric signal after being filtered by the 4-step Carré PSA. The Carré $G(\omega_0)$ peaks at $\omega_0 = 0.5\pi$, so the best carrier for this self-tuning PSA is around $\omega_0 = 0.5\pi$.

We emphasize that, for all possible tuning frequencies within $(0, \pi)$ figure 2 shows that the (S/N) peaks at $\omega_0 = 0.5\pi$, This result is very useful, because now we are certain that the best (S/N) ratio is obtained around $\omega_0 = 0.5\pi$. This result has never been published. This value correct previous publication on this theme [2,3,8].

## 7. Harmonics rejection of the non-linear 4-step Carré algorithm

In Fig. 3 we show the magnitude of the FTF $|H_2(\omega)|$ (Eq. (9)) of the Carré's algorithm for three different carriers: $\omega_0 = 0.5\pi$, $\omega_0 = 0.25\pi$, and $\omega_0 = 0.7\pi$ within $\omega \in [-6\omega_0, 6\omega_0]$.

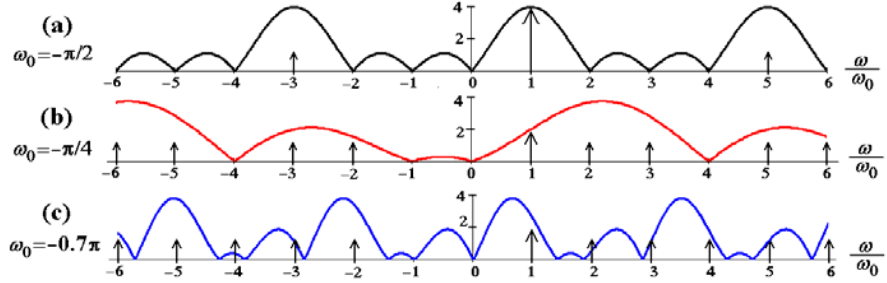

Figure 3. Harmonic analysis of the Carré PSA. The searched analytical signal is at $\omega/\omega_0 = 1.0$. The best harmonic rejection of the Carre PSA is for $\omega_0 = 0.5\pi$. Outside this carrier the harmonics rejection capability degrades substantially.

We see that for $\omega_0 = 0.5\pi$ the Carré PSA is insensitive to harmonics {-6, -5, -4, -2, -1, 2, 3, 4, 6} within the range shown. These include all even harmonics mentioned in [2]. But outside $\omega_0 = 0.5\pi$ (as shown in panels 3(a) and 3(b)) the harmonic distortion increases substantially. All these mean again, that our data must be tuned around $\omega_0 = \pi/2$ for best harmonic and noise rejection. Again, these results on the harmonic rejection of the Carré algorithm have never been published.

## 8. Conclusion

We have analyzed the non-linear 4-steps Carré algorithm in a new perspective. This new perspective uses a slightly different form of it (Eq. (7)) which can be analyzed using linear system theory. The non-linear part estimates the local phase-step $\omega_0(x,y)$, while the linear part estimates $\varphi(x,y)$. Also we demonstrate that the non-linear estimate of $\omega_0(x,y)$ does not play any role in the properties herein studied namely: the frequency response, the noise rejection and the harmonics response. The key of the results presented are contained in the new arrangement proposed of the linear part of the Carré algorithm (Eq. (7)). This new arrangement, however elementary, permits to find explicit formulae for: a) the frequency response, b) the signal-to-noise (S/N) power ratio, and c) the harmonics robustness of the non-linear 4-step Carré PSA. All the explicit mathematical formulae found herein, as indicated at the end of each section, are useful and have never been published.

Finally we comment that the presented technique to analyze the non-linear 4-step Carré algorithm may be extended to higher order self-tuning algorithms. Doing this, and having enough data, you may generate many self-tuning PSAs having any desired properties of spectrum, noise, and harmonic rejection.